\documentclass[twocolumn,english,aps,prl,10pt,superscriptaddress]{revtex4}
\usepackage[T1]{fontenc}
\usepackage{amsmath,graphicx,amssymb,epsfig,babel,dsfont,bm}
\usepackage{mathrsfs}
\usepackage{color}
\usepackage{amsbsy}
\usepackage{lmodern}

\usepackage[bbgreekl]{mathbbol}
\DeclareSymbolFontAlphabet{\mathbbm}{bbold}
\DeclareSymbolFontAlphabet{\mathbb}{AMSb}

\begin{document}

\title{Radiative Corbino effect in nonreciprocal many-body systems}
\author{Ivan Latella}
\email{ilatella@ub.edu}
\affiliation{Departament de F\'{i}sica de la Mat\`{e}ria Condensada, Universitat de Barcelona, Mart\'{i} i Franqu\`{e}s 1, 08028 Barcelona, Spain.}
\affiliation{Institut de Nanociència i Nanotecnologia (IN2UB), 08028 Barcelona, Spain.}
\author{Philippe Ben-Abdallah}
\email{pba@institutoptique.fr}
\affiliation{Laboratoire Charles Fabry, UMR 8501, Institut d'Optique, CNRS, Universit\'{e} Paris-Saclay, 2 Avenue Augustin Fresnel, 91127 Palaiseau Cedex, France.}


\begin{abstract}
When a magnetic field is applied in the perpendicular direction to a metallic disk under the action of a radial bias voltage, a tangential electric current superimposes to the radial current due to the presence of the Lorentz force which acts on electrons. Here we introduce a thermal analog of this Corbino effect in many-body systems made of nonreciprocal bodies which interact by exchanging photons in near-field regime. In systems out of thermal equilibrium with a radial temperature gradient, we demonstrate that the Poynting field in the Corbino geometry is bent in presence of an external magnetic field, giving rise to a tangential heat flux. This thermomagnetic effect could find applications in the field of thermal management and energy conversion at nanoscale.
\end{abstract}
\maketitle

Radiative heat exchange in many-body systems composed of multiple thermal emitters capable of cooperative interactions is an emerging field of research, driven by the unique effects arising from their collective behavior~\cite{Biehs1}. In non-reciprocal many-body systems, the time-reversal symmetry breaking in
Maxwell’s equations leads to a class of peculiar phenomena such as the existence of supercurrents in systems at equilibrium~\cite{Fan}, thermal rectification~\cite{Ott_rect}, a giant magneto-resistance in magneto-optical networks~\cite{Latella2} or the existence of a photon thermal Hall effect~\cite{pba2016}. 
In the present letter we demonstrate that for many-body systems made of magneto-optical objects distributed in a two dimensional circular geometry, the Poynting vector is curled  in presence of a radial temperature gradient under the action of an external magnetic field applied in the direction orthogonal to the disk. This twist of the Poynting field induces a heat flux transversal to the direction of the primary temperature gradient, in a similar way to the Corbino effect~\cite{Corbino,Koch} in electronics.

\begin{figure}[bt]
\centering
\includegraphics[width=\columnwidth]{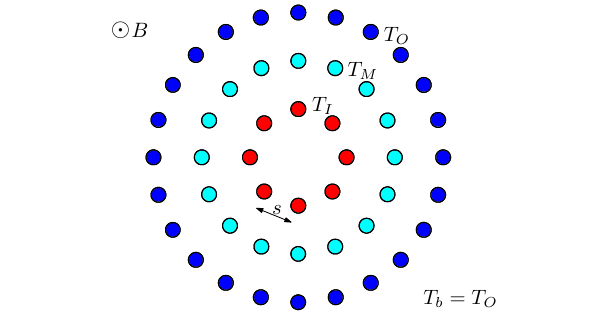}
\caption{Corbino geometry. A disk-like network of magneto-optical particles is subject to a radial temperature gradient in presence of an  orthogonal magnetic field $B$.
Particles in the inner and outer rings have fixed temperatures $T_I$ and $T_O$, respectively, while those in the middle ring are allowed to reach a stationary state characterized by an average temperature $T_M$.
Background thermal radiation is considered at temperature $T_b=T_O$. The particles are spheres of radius $r_p=100\,$nm and $s=5r_p$ specifies the interparticle distance in the inner ring whose radius is $R_I$. The middle and outer rings have radius $R_M=2R_I$ and $R_O=3R_I$, respectively. We consider $N_I=8$, $N_M=16$, and $N_O=24$ particles in the inner, middle, and outer rings, respectively.}
\label{Fig:Sketch}
\end{figure}

To illustrate this effect, we consider the system depicted in Fig.~\ref{Fig:Sketch}, which consists of $N$ spherical particles of radius $r_p$ made of a magneto-optical material, specifically InSb, arranged on a disk in a Corbino-like geometry. Each particle is positioned at $\mathbf{r}_j$ and has temperature $T_j$ for $j=1,\dots,N$. 
A radial temperature gradient is applied to the disk in such a way that particles in the inner ring have temperature $T_I$ and particles in the outer ring have temperature $T_O$.
These particles can exchange electromagnetic energy between them and with the surrounding medium which can be assimilated to a thermal bath at temperature $T_b$, which we set as $T_b=T_O$. A magnetic field of magnitude $B$ is then applied in the direction perpendicular to the disk, so 
the permittivity tensor of these particles at frequency $\omega$ takes the form~\cite{Palik,Moncada}
\begin{equation}
  \bbespilon(\omega)=
  \begin{pmatrix}
  \varepsilon_{1} & -i\varepsilon_{2} & 0\\
  i\varepsilon_{2} & \varepsilon_{1} & 0\\
  0 & 0 & \varepsilon_{3}
  \end{pmatrix}
\label{Eq:permittivity}
\end{equation}
with
\begin{align}
\frac{\varepsilon_{1}(\omega)}{\varepsilon_\infty}&=1+\frac{\omega_L^2-\omega_T^2}{\omega_T^2-\omega^2-i\Gamma\omega}+\frac{\omega_p^2(\omega+i\gamma)}{\omega[\omega_c^2-(\omega+i\gamma)^2]} \label{Eq:permittivity1},\\
\frac{\varepsilon_{2}(\omega)}{\varepsilon_\infty}&=\frac{\omega_p^2\omega_c}{\omega[(\omega+i\gamma)^2-\omega_c^2]}\label{Eq:permittivity2},\\
\frac{\varepsilon_{3}(\omega)}{\varepsilon_\infty}&=1+\frac{\omega_L^2-\omega_T^2}{\omega_T^2-\omega^2-i\Gamma\omega}-\frac{\omega_p^2}{\omega(\omega+i\gamma)}\label{Eq:permittivity3}.
\end{align}
Here, $\varepsilon_\infty=15.7$ is the infinite-frequency dielectric constant, $\omega_L=3.62\times10^{13}\,$rad/s is the longitudinal optical phonon frequency, $\omega_T=3.39\times10^{13}\,$rad/s is the transverse optical phonon frequency, $\omega_p=(\frac{ne^2}{m^*\varepsilon_0\varepsilon_\infty})^{1/2}$ is the plasma frequency of free carriers with density $n=1.36\times10^{19}\,$cm$^{-3}$ and effective mass $m^*=7.29\times10^{-32}\,$kg, $\Gamma=5.65\times10^{11}\,$rad/s is the phonon damping constant, $\gamma=10^{12}\,$rad/s is the free carrier damping constant and $\omega_c(B)=eB/m^*$ is the cyclotron frequency~\cite{Palik,Law}, $e$ being the elementary charge and $\varepsilon_0$ the permittivity of vacuum. Thus, the anisotropic polarizability of the $j$th particle, including the radiative corrections, can be described by the tensor~\cite{Albaladejo}
\begin{equation}
\bbalpha_{j}(\omega)=\left( \mathbb{I}-i\frac{k_0^3}{6\pi} \bbalpha_{0j}\right)^{-1} \bbalpha_{0j}\label{Eq:Polarizability},
\end{equation}
where $\mathbb{I}$ denotes the identity, $ \bbalpha_{0j}$ is the quasistatic polarizability which for a sphere made of a magneto-optical material embedded in vacuum reads 
\begin{equation}
  \bbalpha_{0j}(\omega)=4\pi r_p^3(\bbespilon-\mathbb{I})(\bbespilon+2\mathbb{I})^{-1}\label{Eq:Polarizability2},
\end{equation}
and $k_0=\omega/c$, $c$ being the speed of light in vacuum.

The net power absorbed by a particle in an arbitrary network can be computed using the Landauer formalism for $N$-body systems~\cite{pba2011,Cuevas,Cuevas2}. From this formalism, the non-equilibrium power received by the $j$th particle reads 
\begin{equation}
\mathcal{P}_j=3\int_0^\infty\frac{d\omega}{2\pi}\sum_{k=1}^N \hbar\omega[n(\omega,T_k)-n(\omega,T_b)]\mathcal{T}_{jk}(\omega),
\label{Eq:power}
\end{equation}
where $n(\omega,T)=1/(e^{\hbar\omega/k_B T}-1)$ is the Bose distribution at temperature $T$ and $\mathcal{T}_{jk}(\omega)$ is the energy transmission coefficient between particles $j$ and $k$, $\hbar$ and $k_B$ being the reduced Planck constant and Boltzmann constant, respectively. In the dipolar approximation the transmission coefficients read~\cite{Cuevas,Cuevas2,Ott}
\begin{equation}
\mathcal{T}_{jk}(\omega)= \frac{4}{3}\varepsilon_0\mathrm{Im}\mathrm{Tr}\big[(\mathbb{T}^{-1})_{jk}\bbchi_{k}
(\mathbb{D}\mathbb{T}^{-1})_{jk}^\dagger\big],
\end{equation}
where $(\mathbb{T}^{-1})_{jk}$ denotes the $3\times3$ submatrices of the $3N\times3N$ block matrix $\mathbb{T}^{-1}$, whose inverse $\mathbb{T}$ is defined by the submatrices $\mathbb{T}_{jk}=\delta_{jk}\mathbb{I}-(1-\delta_{jk})k_0^2\bbalpha_j\mathbb{G}_0^{jk}$~\cite{Riccardo}. Here we introduced the particle  susceptibility
\begin{equation}
\bbchi_j=\frac{\bbalpha_j - \bbalpha_j^{\dagger}}{2i}  - k_0^2\bbalpha_j \mathrm{Im}(\mathbb{G}_0^{jj})\bbalpha_j^{\dagger},
\end{equation}
the block matrix $\mathbb{D}$ defined by the submatrices $\mathbb{D}_{jk}=\frac{k_0^2}{\varepsilon_0}\mathbb{G}_0^{jk}$,
and the notation $\mathbb{G}_0^{jk}\equiv\mathbb{G}_0^\mathrm{EE}(\mathbf{r}_j,\mathbf{r}_k,\omega)$ to specify the electric-electric Green tensor in vacuum evaluated at the positions of particles $j$ and $k$~\cite{Riccardo}. Furthermore, in our Corbino configuration, particles in the inner and outer rings have fixed temperatures $T_I$ and $T_O$, respectively. The remaining particles are not thermostated and are allowed to reach stationary temperatures $T_j^\mathrm{st}$ for which the net power they absorbe vanishes, $\mathcal{P}_j=0$ for $j$ in the middle ring. Due to the discrete nature of the system, the radial symmetry is only approximate and particles along the middle ring may possess different neighbor configurations. This makes that the temperatures $T_j^\mathrm{st}$ may be slightly different (although $\mathcal{P}_j=0$ for all of them). Here we characterize the temperature $T_M$ of the middle ring by taking an average over $T_j^\mathrm{st}$.

To investigate how nonreciprocity and the breaking of time-reversal symmetry in Maxwell’s equations impact the energy exchange between particles, we examine the behavior of the mean Poynting vector $\langle\mathbf{S}(\mathbf{r})\rangle$ at points $\mathbf{r}$ in the system. Since $\langle\mathbf{S}(\mathbf{r})\rangle$ characterizes energy fluxes, the absorbed power (\ref{Eq:power}) is given by $\mathcal{P}_j=-\int_{\Sigma_j} \langle\mathbf{S}(\mathbf{r})\rangle\cdot d\bm{\Sigma}_j$, where $\Sigma_j$ is a surface enclosing the $j$th particle.
By definition, the statistical average of the Poynting vector can be written as $\langle\mathbf{S}(\mathbf{r})\rangle=\int_0^\infty\frac{d\omega}{2\pi}\tilde{\mathbf{S}}(\mathbf{r},\omega)$ with the stationary spectrum given by
\begin{equation}
 \tilde{\mathbf{S}}(\mathbf{r},\omega) \equiv 2 \mathrm{Re}\langle\mathbf{E}(\mathbf{r},\omega)\times \mathbf{H}^*(\mathbf{r},\omega)\rangle,
\label{Poynting}
\end{equation}
in which the Fourier components of the electric and magnetic fields $\mathbf{E}(\mathbf{r},\omega)$ and $\mathbf{H}(\mathbf{r},\omega)$, respectively, can be related to local fluctuating dipoles $\bar{\mathbf{p}}_j$ according to
\begin{align}
 \mathbf{E}(\mathbf{r},\omega)&=\mathbf{E}^b(\mathbf{r},\omega)+\frac{k_0^2}{\varepsilon_0} \sum_j\mathbb{G}^\mathrm{EE}(\mathbf{r},\mathbf{r}_j,\omega)\bar{\mathbf{p}}_j,\label{Electric}\\
 \mathbf{H}(\mathbf{r},\omega)&=\mathbf{H}^b(\mathbf{r},\omega)+\frac{k_0^2}{\varepsilon_0} \sum_j\mathbb{G}^\mathrm{HE}(\mathbf{r},\mathbf{r}_j,\omega)\bar{\mathbf{p}}_j.\label{Magnetic}
\end{align}
where $\mathbf{E}^b$ and $\mathbf{H}^b$ denote the electric and magnetic field contributions coming from the thermal bath, while the second terms in these expressions set the fields scattered by the dipoles, $\mathbb{G}^\mathrm{EE}$ and $\mathbb{G}^\mathrm{HE}$ being the full electric and magnetic Green tensors, respectively. These Green tensors read~\cite{supplemental}
\begin{align}
\mathbb{G}^\mathrm{EE}(\mathbf{r},\mathbf{r}_k,\omega)&=\sum_j\mathbb{G}_0^\mathrm{EE}(\mathbf{r},\mathbf{r}_j,\omega)(\mathbb{T}^{-1})_{jk},\\
\mathbb{G}^\mathrm{HE}(\mathbf{r},\mathbf{r}_k,\omega)&=\sum_j\mathbb{G}_0^\mathrm{HE}(\mathbf{r},\mathbf{r}_j,\omega)(\mathbb{T}^{-1})_{jk},
\end{align}
$\mathbb{G}_0^\mathrm{HE}(\mathbf{r}_j,\mathbf{r}_k,\omega)$ being the magnetic-electric Green tensor in vacuum. For simplicity, below we omit making explicit the dependence on frequency.

\begin{figure}
\centering
\includegraphics[width=\columnwidth]{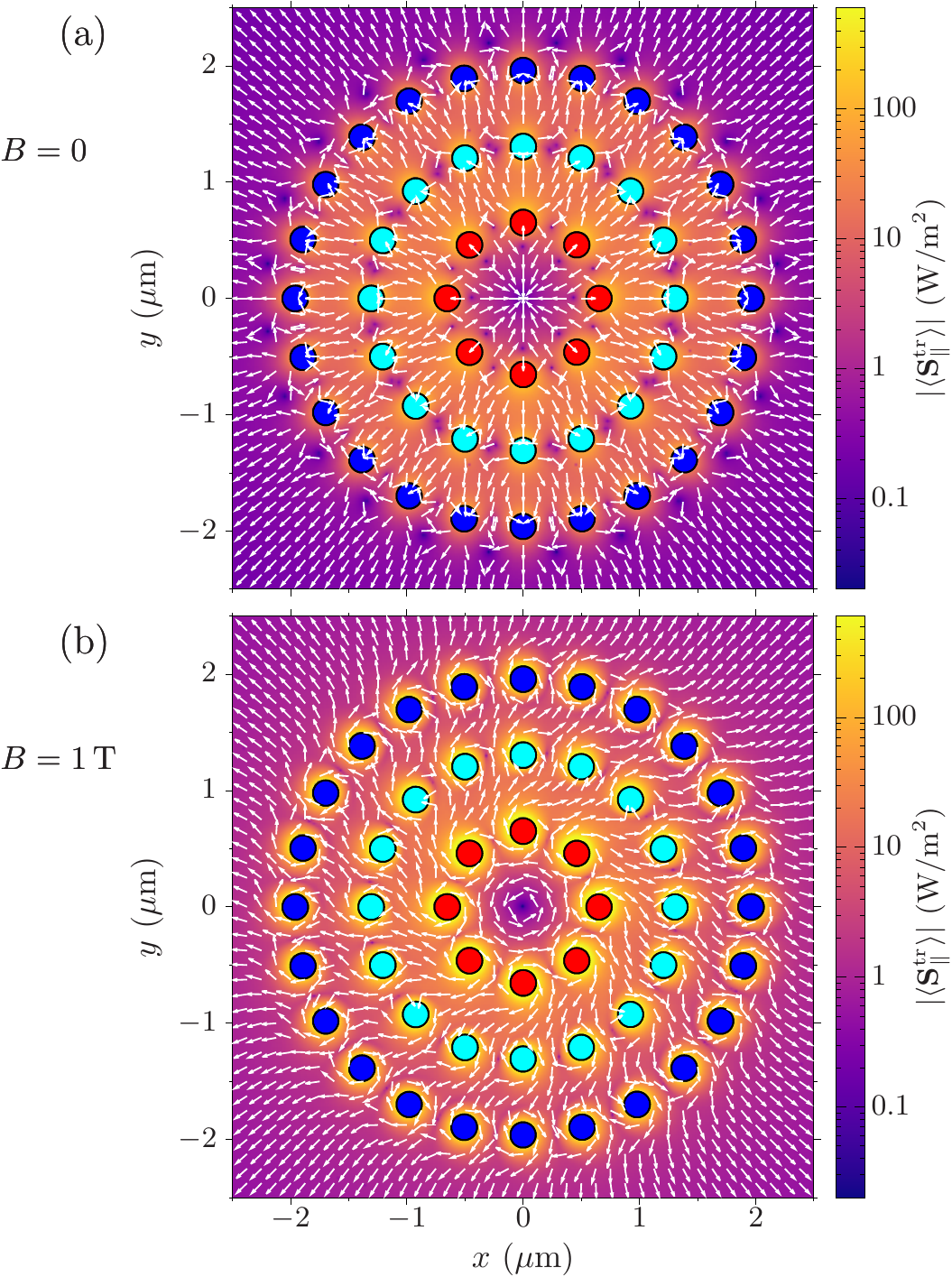}
\caption{Radiative Corbino effect characterized by the mean Poynting vector in the plane containing InSb particles. (a) In absence of the magnetic field, the energy flux exhibits a radial trend following the discrete distribution of the particles. (b) For an applied magnetic field, the Poynting vector bends due to the magneto-optical response of the particles. The radius of the particles is $r_p=100\,$nm, the interparticle distance in the inner ring is $s=5r_p$, and the temperatures of the inner and outer rings are $T_I=350\,$K and $T_O=300\,$K, respectively.}
\label{Fig:Poynting}
\end{figure}

\begin{figure}
\centering
\includegraphics[width=\columnwidth]{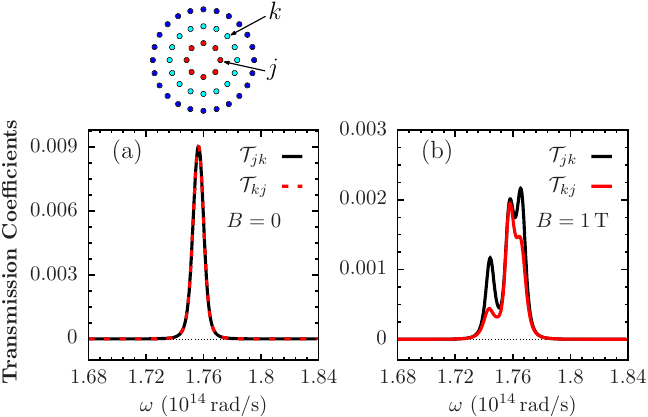}
\caption{Transmission coefficients between two InSb particles for the configuration considered in Fig.~\ref{Fig:Poynting}. (a) The coefficients coincide at zero magnetic field. (b) The coefficients are different due to nonreciprocity at nonzero applied field.}
\label{Fig:Trans_coeff}
\end{figure}

The dipole moment $\bar{\mathbf{p}}_j=\mathbf{p}^f_j+\varepsilon_0\bbalpha_j\mathbf{E}^b(\mathbf{r}_j)$ has a fluctuating contribution $\mathbf{p}^f_j$ arising from intrinsic fluctuations in the particle and an induced contribution $\varepsilon_0\bbalpha_j\mathbf{E}^b(\mathbf{r}_j)$ coming from the fluctuating background electric field. Taking this into account, the averaged Poynting vector can be computed by means of the fluctuation-dissipation theorem~\cite{Riccardo,Callen}
\begin{equation}
\langle \mathbf{p}_{j}^f\otimes\mathbf{p}_{k}^{f*}\rangle =\frac{2\varepsilon_0}{\omega}\Theta_j\bbchi_{j} \delta_{jk}
\label{FDT_dipoles} 
\end{equation}
and~\cite{Agarwal,Eckhardt}
\begin{align}
\langle \mathbf{E}^b(\mathbf{r}_j)\otimes \mathbf{E}^{b*}(\mathbf{r}_k)\rangle
&=\frac{2 k_0}{\varepsilon_0c}\Theta_b\mathrm{Im}[\mathbb{G}_0^\mathrm{EE}(\mathbf{r}_j,\mathbf{r}_k)] \label{FDT_E_bath_field}, \\
\langle \mathbf{E}^b(\mathbf{r}_j)\otimes \mathbf{H}^{b*}(\mathbf{r}_k)\rangle
&=i\frac{2 k_0}{\varepsilon_0c}\Theta_b\mathrm{Re}[\mathbb{G}_0^\mathrm{HE}(\mathbf{r}_j,\mathbf{r}_k)] \label{FDT_E_H_bath_field},
\end{align}
where $\Theta_j=\hbar\omega[n(T_j)+\frac{1}{2}]$ and $\Theta_b=\hbar\omega[n(T_b)+\frac{1}{2}]$.
Furthermore, from Eq.~(\ref{Poynting}), the Poynting vector spectrum can be decomposed as~\cite{Ott} $\tilde{\mathbf{S}}(\mathbf{r})=\tilde{\mathbf{S}}^\mathrm{tr}(\mathbf{r})+\tilde{\mathbf{S}}^0(\mathbf{r})$ with the components of these vector fields written as
\begin{align}
\tilde{S}^\mathrm{tr}_{\alpha}(\mathbf{r})&=2\mathrm{Re}\sum_{\beta,\gamma}\epsilon_{\alpha\beta\gamma}[\mathbb{M}^\mathrm{tr}(\mathbf{r})]_{\beta\gamma},\label{poynting_components_tr}\\
\tilde{S}^0_{\alpha}(\mathbf{r})&=2\mathrm{Re}\sum_{\beta,\gamma}\epsilon_{\alpha\beta\gamma}[\mathbb{M}^0(\mathbf{r})]_{\beta\gamma},\label{poynting_components_0}
\end{align}
where $\epsilon_{\alpha\beta\gamma}$ is the Levi-Civita symbol for $\alpha,\beta,\gamma=x,y,z$. Here we have introduced the matrices
\begin{align}
\mathbb{M}^\mathrm{tr}(\mathbf{r})
&=\frac{2k_0^3}{\varepsilon_0c}
\sum_{j}\left(\Theta_j-\Theta_b\right)\mathbb{G}^\mathrm{EE}(\mathbf{r},\mathbf{r}_j)
\bbchi_{j}
\mathbb{G}^\mathrm{HE\dagger}(\mathbf{r},\mathbf{r}_j)
\label{Mtr}
\end{align}
and
\begin{equation}
\begin{split}
\mathbb{M}^0(\mathbf{r})&=\frac{k_0^3 \Theta_b}{i\varepsilon_0c}\sum_{j}\Big\{\mathbb{G}^\mathrm{EE}(\mathbf{r},\mathbf{r}_j) \bbalpha_j\mathbb{G}_0^\mathrm{HE}(\mathbf{r},\mathbf{r}_j)\\
&-\left[\mathbb{G}^\mathrm{HE}(\mathbf{r},\mathbf{r}_j)\bbalpha_j\mathbb{G}_0^\mathrm{EE}(\mathbf{r},\mathbf{r}_j)\right]^\dagger\Big\} 
\end{split}
\label{M0}
\end{equation}
which are obtained from the electric-magnetic correlation matrix according to $\langle\mathbf{E}(\mathbf{r})\otimes \mathbf{H}^*(\mathbf{r})\rangle=\mathbb{M}^\mathrm{tr}(\mathbf{r})+\mathbb{M}^0(\mathbf{r})$. Notice that $\mathbb{M}^\mathrm{tr}(\mathbf{r})$ vanishes when the system is in thermal equilibrium, but for general nonreciprocal systems $\mathbb{M}^0(\mathbf{r})$ does not; in particular, the latter depends on the temperature of the background and on zero-point fluctuations. However, as anticipated in Ref.~\cite{Ott} without proof, only $\tilde{\mathbf{S}}^\mathrm{tr}(\mathbf{r})$ contributes to the heat transfer. In the supplemental material~\cite{supplemental} we give a general demonstration on the fact that $\nabla\cdot\tilde{\mathbf{S}}^0(\mathbf{r})=0$, and therefore the absorbed power~(\ref{Eq:power}) is actually given by $\mathcal{P}_j=-\int_{\Sigma_j} \langle\mathbf{S}^\mathrm{tr}(\mathbf{r})\rangle\cdot d\bm{\Sigma}$, where 
$\langle\mathbf{S}^\mathrm{tr}(\mathbf{r})\rangle=\int_0^\infty\frac{d\omega}{2\pi}\tilde{\mathbf{S}}^\mathrm{tr}(\mathbf{r},\omega)$.

In Fig.~\ref{Fig:Poynting}, we show the in-plane component of the Poynting vector $\langle\mathbf{S}^\mathrm{tr}(\mathbf{r})\rangle$ inside the Corbino geometry in presence and without magnetic field. When $B=0$  all particles are isotropic and the system is reciprocal. In this case, represented in Fig.~\ref{Fig:Poynting}(a), there is no symmetry breaking in the system so that there is no tangential flux through the network. On the contrary, when a magnetic field is applied orthogonally to the disk, particles become anisotropic and the symmetry of the system is broken, as shown in Fig.~\ref{Fig:Poynting}(b). The induced circulation of heat flux  in the tangential direction is a generalization of the effect shown in~\cite{Ott2} for a single particle.

\begin{figure*}[!t]
\centering
\includegraphics[width=\linewidth]{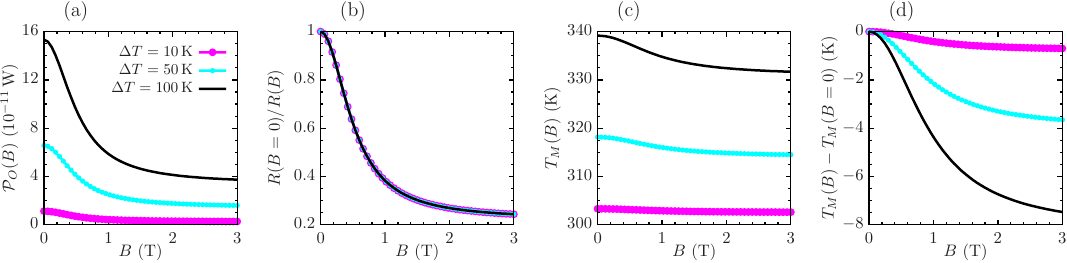}
\caption{(a) Power absorbed by the outer ring in the considered Corbino geometry as a function of the applied magnetic field for different temperature differences $\Delta T=T_I-T_O$ with $T_O=T_b=300\,$K. (b) Corbino thermal magneto-resistance and (c) stationary temperature of the middle ring as a function of the magnetic field for the different values of $\Delta T$. (d) Variation of the temperature of the middle ring as a function of the magnetic field corresponding to (c).}
\label{Fig:Resistance}
\end{figure*}

The breaking of symmetry responsible for this effect is illustrated in Fig.~\ref{Fig:Trans_coeff}, where we represent the energy transmission coefficients $\mathcal{T}_{jk}$ and $\mathcal{T}_{kj}$ between two particles $j$ and $k$ under the action of an external magnetic field. Moreover, as it can be seen in this figure, the magnitude of the transmission coefficients decreases when the magnetic field is applied, showing a reduction of the energy exchange between the particles. This reduction of  heat flux carried by the thermal photons is directly related to the occurrence of a thermal magneto-resistance $R(B)$ inside the system, as previously highlighted in~\cite{Latella2}. Noting that the particles in the outer ring at $T_O$ exchange thermal radiation with those in the inner ring at $T_I$ (the environmental temperature is $T_b=T_O$), the thermal magneto-resistance can be computed as
\begin{equation}
R(B)= \frac{\Delta T}{\mathcal{P}_O(B)},
\end{equation}
where $\Delta T= T_I-T_O$ and $\mathcal{P}_O(B)=\sum_{j_O}\mathcal{P}_j$, $j_O$ labeling particles in the outer ring.  The total power absorbed by the outer ring $\mathcal{P}_O(B)$ is shown in Fig.~\ref{Fig:Resistance}(a) for the previously considered geometrical arrangement by taking different values of $\Delta T$. It can be seen that $\mathcal{P}_O(B)$ decreases as the field increases, so that $R(B)$ grows accordingly. Moreover, the ratio $R(B=0)/R(B)$ as represented in Fig.~\ref{Fig:Resistance}(b) with respect to the magnitude of $B$ is almost constant for the different temperature differences we have considered, in agreement with the fact that $\mathcal{P}_O(B)$ is a linear function  of $\Delta T$ when $\Delta T$ remains relatively small. In addition, the magneto-resistance induces a drop in the stationary temperature $T_M$ of the middle ring as highlighted in Figs.~\ref{Fig:Resistance}(c) and \ref{Fig:Resistance}(d). It is interesting to note that this temperature can undergoes a variation of several degrees with changes of the magnetic field around $1\,$T. These results show that this Corbino effect could be used to modulate the inner temperature of a many-body system in order to locally control heat exchange through shuttling~\cite{Latella3} and to convert heat into electric current by coupling these systems with pyrolectric devices~\cite{Latella4}.
 
We have demonstrated the existence of a thermal analog of the Corbino effect due to the interaction of thermal photons in magneto-optical many-body systems. We have shown, in these systems, that the Poynting vector is curled in presence of a radial temperature gradient under the action of an external magnetic field. This twist of the Poynting field induces a heat flux transversal to the direction of the primary temperature gradient. In addition to its potential for thermal management and energy conversion, this effect could give rise to a tangential force on each component of the system, resulting in an overall torque. This opens up possibilities for developing thermal ratchets capable of converting local heat sources into mechanical work.

\begin{acknowledgments}
We thank fruitful discussions with M. Antezza and S.-A. Biehs.
I.L. acknowledges financial support from the Spanish Government through Grant No. PID2021-126570NBI00 (MICIU/FEDER, UE). P.B.-A. acknowledges financial support from the French Agence Nationale de la Recherche (ANR), under grant ANR-21-CE30-0030 (NBODHEAT).
\end{acknowledgments}

\end{document}